\input amssym.def
\input amssym
\newcommand{\Beq}{\begin{equation}}
\newcommand{\Eeq}{\end{equation}}
\newcommand{\Beqa}{\begin{eqnarray}}
\newcommand{\Eeqa}{\end{eqnarray}}
\newcommand{\End}{\nonumber\\}
\newcommand{\Lba}{\bigl(}
\newcommand{\Lbb}{\Bigl(}

\newcommand{\Rba}{\bigr)}
\newcommand{\Rbb}{\Bigr)}

\newcommand{\Lsa}{\bigl[}
\newcommand{\Rsa}{\bigr]}

\newcommand{\Seto}{\{}
\newcommand{\Setc}{\}}

\newcommand{\Half}{{\textstyle\frac{1}{2}}}

\newcommand{\Smat}[4]{\left(\begin{array}{cc} #1 & #2 \\
#3 & #4 \end{array}\right)}
\newcommand{\La}{\Lambda}
\newcommand{\Id}{I}
\newcommand{\Em}{E_M}
\newcommand{\Bare}{{\Bbb C}_{S}}
\newcommand{\Comp}{{\Bbb C}}
\newcommand{\Ho}[1]{{\cal O}(#1)}

\newcommand{\Codd}{\Bare{}_1}
\newcommand{\Ceven}{\Bare{}_0}
\newcommand{\Sc}{{\cal SC}}
\newcommand{\Hom}{\Ho{\Bod M} }
\newcommand{\Sb}{{\scriptstyle{K}}}
\newcommand{\Sqk}{\sqrt{\Sb}}
\newcommand{\Spim}{\Gamma(\Sqk) }
\newcommand{\Bod}[1]{#1{}_{\{\emptyset\}}}
\newcommand{\Ge}[1]{\nu\Ind{#1}}
\newcommand{\Ind}[1]{{}_{\{#1\}}}
\newcommand{\Indd}[1]{{}_{#1}}
\newcommand{\Lev}[1]{{}_{[[#1]]}}
\newcommand{\Ep}{\epsilon}
\newcommand{\One}{{\bf 1}}
\newcommand{\al}{\alpha}
\newcommand{\be}{\beta}
\newcommand{\ga}{\gamma}
\newcommand{\zal}{z_{\al}}
\newcommand{\zbal}{\Bod{z_{\al}}}
\newcommand{\zeal}{\zeta_{\al}}
\newcommand{\zbe}{z_{\be}}
\newcommand{\zbbe}{\Bod{z_{\be}}}
\newcommand{\zebe}{\zeta_{\be}}
\newcommand{\zga}{z_{\ga}}
\newcommand{\zega}{\zeta_{\ga}}
\newcommand{\Dal}{D_{\al}}
\newcommand{\Dbe}{D_{\be}}

\newcommand{\Ual}{U_{\al}}
\newcommand{\Ube}{U_{\be}}

\newcommand{\Val}{V_{\al}}
\newcommand{\Vbe}{V_{\be}}
\newcommand{\Vga}{V_{\ga}}
\newcommand{\gal}{g_{\al}}

\newcommand{\pal}{p_{\al}}

\newcommand{\ral}{r_{\al}}
\newcommand{\rbe}{r_{\be}}

\newcommand{\pial}{\pi_{\al}}

\newcommand{\pialz}{\pi_{\al}(\zal)}

\newcommand{\Ab}[1]{#1{}_{\al\be}}
\newcommand{\Mab}{\Ab M}
\newcommand{\Mabzz}{\Ab M(\zbe,\zebe)}
\newcommand{\Fab}{\Ab f}
\newcommand{\Psab}{\Ab{\psi}}

\newcommand{\Fabz}{\Ab f(\zbe)}
\newcommand{\Psabz}{\Ab{\psi}(\zbe)}
\newcommand{\Pspabz}{\Ab{\psi'{}}(\zbe)}
\newcommand {\Fpabz}{\Ab{f'{}}(\zbe)}
\newcommand{\Bg}[1]{#1{}_{\be\ga}}

\newcommand{\Fbg}{\Bg f}
\newcommand{\Psbg}{\Bg{\psi}}

\newcommand{\Psbgz}{\Bg{\psi}(\zga)}

\newcommand{\Ag}[1]{#1{}_{\al\ga}}

\newcommand{\Fag}{\Ag f}
\newcommand{\Psag}{\Ag{\psi}}

\newcommand{\Psagz}{\Ag{\psi}(\zga)}

\newcommand{\Nab}{\Ab{{\cal N}}}
\newcommand{\Nabzz}{\Ab{{\cal N}}(\zbe,\zebe)}

\newcommand{\Nbgzz}{\Bg{{\cal N}}(\zga,\zega)}

\newcommand{\Nagzz}{\Ag{{\cal N}}(\zga,\zega)}
\newcommand{\Fbab}{\Bod{\Ab f}}

\newcommand{\Fbag}{\Bod{\Ag f}}

\newcommand{\Fbbg}{\Bod{\Bg f}}

\newcommand{\Fbabzb}{\Bod{\Ab f}(\Bod{\zbe})}

\newcommand{\Lab}{L_{\al\be}}
\newcommand{\Lbg}{L_{\be\ga}}
\newcommand{\Lag}{L_{\al\ga}}
\newcommand{\TTab}{T_{\al\be}}
\newcommand{\TTag}{T_{\al\ga}}
\newcommand{\TTbg}{T_{\be\ga}}
\newcommand{\Aab}{A_{\al\be}}
\newcommand{\Aag}{A_{\al\ga}}
\newcommand{\Abg}{A_{\be\ga}}
\newcommand{\Ffab}{F_{\al\be}\Indd{ijk}}
\newcommand{\Ffag}{F_{\al\ga}\Indd{ijk}}
\newcommand{\Ffbg}{F_{\be\ga}\Indd{ijk}}
\newcommand{\Kab}{K_{\al\be}\Indd{ijk}}
\newcommand{\Kag}{K_{\al\ga}\Indd{ijk}}
\newcommand{\Kbg}{K_{\be\ga}\Indd{ijk}}
\newcommand{\Uabg}{U_{\alpha\beta\gamma}}
\newcommand{\Vec}[2]{\left( \begin{array}{c}
#1\\#2 \end{array}\right)}
\newcommand{\Svec}[2]{\left\{\left( \begin{array}{c}
#1\\#2 \end{array}\right)\right\}}

\newcommand{\Vecal}{\Vec{g_\alpha}{\gamma_{\alpha}}}
\newcommand{\Vecbe}{\Vec{g_\beta}{\gamma_{\beta}}}

\newcommand{\Coh}[1]{H\sp{#1}(\Bod{M},{\cal O})}
\newcommand{\Cha}[1]{C\sp{#1}(\Bod{M},{\cal O})}
\newcommand{\sigal}{\sigma_{\alpha}}
\newcommand{\sigbe}{\sigma_{\beta}}

\newcommand{\Diff}[2]{\frac{d\,#1}{dz_{#2}}}
\newcommand{\Sigt}{\tilde{\sigma}}
\newcommand{\Lr}{\leftrightarrow}
\newtheorem{Def}{Definition}
\newtheorem{The}{Theorem}
\title{New Fields on Super Riemann Surfaces}
\author{Alice Rogers\thanks{Research supported by a Royal Society
University Research Fellowship}
\ and
Mark Langer\\
Department of Mathematics\\
King's College\\
Strand\\
London WC2R 2LS}
\date{25th May 1994\\ \ \\kcl-th-94-07, hep-th/9405177}
\documentstyle[12pt]{article}
\begin{document}
\bibliographystyle{plain}
\maketitle
\begin{abstract}
A new $(1,1)$-dimensional super vector bundle which exists on
any super Riemann surface is described. Cross-sections of this bundle
provide a new class of fields on a super Riemann surface which closely
resemble holomorphic functions on a super Riemann surface, but which (in
contrast to the case with holomorphic functions) form spaces which have
a well defined dimension which does not change as odd moduli become
non-zero.
\end{abstract}
Super Riemann surfaces are $(1,1)$-dimensional holomorphic complex
supermanifolds which have interesting mathematical features and have
been intensively studied because of their use in a very elegant and
effective
approach to the Polyakov quantization of the spinning string. In this
approach, which has been developed by a number of authors,
for instance in the works of Baranov, Manin, Frolov and Schwarz
\cite{BarMan}, Baranov and Schwarz \cite{BarSch1}, Belavin and Knizhnik
\cite{BelKni}, Rosly, Schwarz and Voronov \cite{RosSchVor2} and Voronov
\cite{Vorono}, many classical techniques of algebraic geometry are
generalised to the super setting. However these methods
depend on certain spaces of fields on super Riemann surfaces having a
super vector space structure (that is, a free module structure) which
they do not in fact possess. Not only do
these spaces lack this structure, but also the nature of these spaces
may vary as one moves around the moduli space of super Riemann surfaces
corresponding to fixed genus and spin structure on the underlying
Riemann
surface. This difficulty, which undermines the analysis used in the
quantization method, has been recognised for some time - it seems to
have been first mentioned in the literature by Giddings and Nelson
\cite{GidNel}, quoting comments of Witten, and then extensively analysed
by Hodgkin \cite{Hodgki}.

The purpose of this paper is to describe a modification of the notion of
function, which is very natural in the context of a super Riemann
surface, and which allows one to construct a family of fields with the
desired super vector space structure. As well as suggesting a means by
which the Polyakov quantization procedure might be made valid, this new
class of functions may make possible further developments in the general
theory of super Riemann surfaces. The construction is based on a
$(1,1)$-dimensional super vector bundle which is shown to be canonically
defined on any super Riemann surface. It makes use of the superconformal
structure of the super Riemann surface, in contrast to the functions
usually considered which only use the superanalytic structure.

The structure of a super Riemann surface
first appears in the work of Howe \cite{Howe} on the superspace
formulation of $2$-dimensional superconformal gravity; here the
construction is in terms of a real $(2,2)$-dimensional supermanifold.
Subsequently this same structure was formulated very elegantly in terms
of $(1,1)$-dimensional complex co-ordinates by Baranov and Schwarz
\cite{BarSch1} and by Friedan \cite{Frieda}. Much of the basic theory of
super Riemann surfaces was developed in detail by Crane and Rabin
\cite{CraRab}, and their notation and terminology is largely used here.

Various equivalent definitions of the notion of super Riemann surface
have been given; we use here an approach based on local co-ordinates
because it is most appropriate for the explicit cohomological
calculations which follow. Briefly, a super Riemann surface is a
$(1,1)$-dimensional complex
holomorphic supermanifold $M$ which satisfies the following additional
condition: there must exist a covering of $M$ by co-ordinate
neighbourhoods $(\Val | \al \in \La)$ with corresponding local
co-ordinates $(\zal,\zeal)$ such that on overlapping co-ordinate
neighbourhoods $\Val$, $\Vbe$  the differential operators $\Dal =
\partial/\partial\zeal + \zeal\partial /\partial\zal$ and $\Dbe =
\partial/\partial\zebe +\zebe\partial/\partial\zbe$ are related by
\Beq
\Dbe = \Mabzz \Dal.
\Eeq
It follows directly from this definition first that
\Beq
\Mabzz = \Dbe \zeal(\zbe,\zebe)
\Eeq
and second that the expression for the co-ordinates $(\zal,\zeal)$ in
terms of the co-ordinates $(\zbe,\zebe)$ takes the form
\Beqa
\zal &=& \Fabz + \zebe \Psabz \sqrt{\Fpabz}\End
\zeal &=& \Psabz + \zebe\sqrt{\Fpabz+\Psabz\Pspabz},
\label{TFeq}
\Eeqa
where the function $\Fab$ is an even holomorphic function and
the function $\Psab$ is an odd holomorphic function. The Grassmann
algebra $\Bare$ on which the super Riemann surface $M$ is
modelled (and in which all functions take values) is here taken to be
the infinite-dimensional complex Grassmann algebra with
generators $1,\Ge1,\Ge2,\dots$ so that a typical element $C$ of $\Bare$
may be expressed as
\Beqa
C = \Bod{C} + \sum_i C\Ind{i} \Ge i + \sum_{i<j}
C\Ind{ij}\Ge i\Ge j +
\dots
\Eeqa
where the coefficients $\Bod C,C\Ind{i},C\Ind{ij}$ etc are
complex numbers. The even and odd parts of $\Bare$ are denoted $\Ceven$
and $\Codd$ respectively. In the proof of the main theorem of this paper
we will
use the $Z$-grading of $\Bare$,
\Beq
\Bare = \Bare\Lev0 \oplus \Bare\Lev1 \oplus \Bare\Lev2 \oplus\dots ,
\Eeq
where $\Bare\Lev{r}$ contains terms of level $r$ in the Grassmann
generators, that is, linear combinations of exactly $r$ anticommuting
generators.  A useful mapping is the mapping $\Ep:\Bare \to \Comp$ which
is defined by
\Beq
\Ep C = \Bod C.
\Eeq
This map
allows one to construct the underlying Riemann surface $\Bod M$
of $M$ as a Riemann surface with co-ordinate neighbourhoods $\Seto
\Ual|\al \in \La \Setc$, where  $\Ual= \epsilon(\Val)$, and local
co-ordinates $\zbal$ which transform
by
\Beq
\zbal = \Fbabzb
\Eeq
where the functions $\Fbab$ are holomorphic functions on $\Ual \cap
\Ube$ such that $\Fbab(\zbbe) = \Ep(\Fab(\zbe))$.
The super Riemann surface $M$ also determines a choice of sign for the
square roots $\sqrt{\Fpabz}$ and hence a spin structure on $\Bod M$. The
corresponding line bundle of spinors is denoted $\Sqk$.

Super Riemann surfaces on which the functions $\Psab$ (which partly
determine
the super Riemann surface structure) all vanish are known as split
super Riemann
surfaces. On such surfaces the local co-ordinate representatives of a
superholomorphic function take the form
\Beq
g_{\al}(\zal,\zeal) = p + \zeal\pialz
\Eeq
where $p$ is a constant element of $\Ceven$ and $\pial$ are local
representatives of a $\Codd$-valued
section of the spin bundle $\Sqk$. Thus the space of all such functions
has the super vector space structure $\Ho T = (\Ceven \otimes \Comp)
\oplus (\Codd \otimes \Spim)$. However (although such structure is
actually required in the superholomorphic quantization of the spinning
string) the space of super holomorphic functions does not take this form
on an arbitrary super Riemann surface. The simplest counterexample
\cite{GidNel,Hodgki} is the non-split super torus with odd modulus
$\psi$. Here the most general even holomorphic
function which can be defined has the local form
\Beq
g(z,\zeta) = c + \gamma\zeta
\Eeq
where $c$ is an arbitrary even constant and $\gamma$ is an
odd
constant such that $\psi\gamma=0$. The set $\Ho T$ of such functions
does not have the structure of a super vector space - that is, there do
not exist complex vector spaces $V_0$ and $V_1$ such that
$$
\Ho T \cong (\Ceven \otimes V_0) \oplus (\Codd \otimes V_1).
$$
(In more technical terms, a super vector space is the even
part of a free graded $\Bare$-module.) The super vector space structure
is important because it guarantees the existence of bases and dimension,
which do not
always exist for a module over a Grassmann algebra.

The starting point of the constructions in this paper is the
observation that, whenever $\Val \cap \Vbe \cap \Vga$ is not empty,
the three functions $\Psab,\Psbg$ and $\Psag$ satisfy the relation
\Beq
\Psagz = \Psabz + \Mabzz \Psbgz
\label{PSeq}
\Eeq
as can be shown using equations (\ref{TFeq}).
Then, if for each $\al,\be$ in $\La$ such that $\Val \cap
\Vbe$ is
not empty one defines the $(1,1) \times (1,1)$ supermatrix
functions
$\Nab$ on $\Val \cap \Vbe$ by
\Beq
\Nabzz = \Smat{1}{\Mabzz\sp{-1}\Psabz}{0}{\Mabzz\sp{-1}},
\label{TRANSeq}\Eeq
three crucial results follow. First, for all $\al$ in $\La$,
\Beq
{\cal N}_{\al\al} = \Id
\Eeq
where $\Id$ denotes the identity matrix, secondly for all
$\al$,$\be$ in
$\La$ such that $\Val \cap \Vbe$ is not empty
\Beq
{\cal N}_{\be\al}(\zal,\zeal) = \Nabzz\sp{-1}
\Eeq
and thirdly for all $\al$,$\be$ and $\ga$ such that $\Val
\cap \Vbe \cap
\Vga$ is not empty
\Beq
\Nagzz = \Nabzz \Nbgzz.
\Eeq
This shows that the collection of functions $\Seto \Nab:\Val
\cap \Vbe \to GL(1,1|\Comp) \Setc$ defines a $(1,1)$-dimensional complex
super vector bundle $\Em$ over the super Riemann surface $M$. Equipped
with this bundle, the key definition of a superconformal function can
now be given. Roughly speaking, the idea is that a superconformal
function should be a cross-section of $\Em$ whose local representatives
have the form
\Beq
\Vec{g_{\alpha}}{D_{\alpha}g_{\alpha}}.
\label{SIMPeq}\Eeq
(The work of Baranov and Schwarz \cite{BarSch1} on zeroes and poles on
super Riemann surfaces demonstrates the
importance of the pair $(g,Dg)$.) In fact this simple
definition of a superconformal function is slightly too restrictive,
the odd component has to have some further freedom, (although it can be
shown that co-ordinates can always be chosen in which a superconformal
function does take the simple form (\ref{SIMPeq})).
\begin{Def}
A superconformal function $G$ on the super Riemann surface $M$ is a
cross-section of $\Em$ whose local representatives $\Vecal$ are
holomorphic and satisfy
\Beq
\gamma_{\alpha}(\zal,\zeal) = D_{\alpha}(g_{\alpha}(\zal,\zeal) +
r_{\alpha}(\zal))
\label{SCeq}\Eeq
for some holomorphic function $\ral$ on $\Val$.
\end{Def}
The main theorem of this paper will now be established. This theorem
shows that the space of superconformal functions on a super Riemann
surface has the desired super vector space structure.
\begin{The}
The space $\Sc(M)$ of superconformal functions on the super
Riemann surface $M$ has the structure
\Beq
\Sc(M)  \cong (\Ceven \otimes \Hom) \oplus (\Codd \otimes \Spim).
\Eeq
\end{The}
Proof

First note that if the vectors
$$
\Vec{g_{\alpha}}{\gamma_{\alpha}} \quad \alpha \in {\Lambda}
$$
are the local representatives of a superconformal function then the
expansions of $g$ and $\gamma$ in terms of $\zal$ and $\zeal$ must (by
equation (\ref{SCeq})) take the form
\Beq
\Vec{g_\alpha}{\gamma_{\alpha}}=\Vec{p_{\alpha}(\zal) +
\zeal\sigma_{\alpha} (\zal)}{\sigma_{\alpha}(\zal) + \zeal
(\pal'(\zal)+r_{\alpha}'(\zal))},
\Eeq
where $\pal,\sigma_{\alpha}$ and $r_{\alpha}$ are holomorphic functions.

Substitution into the transformation law derived from (\ref{TRANSeq})
(inverted, because it leads to
much simpler equations) gives
\Beq
\Vecbe =\Smat{1}{-\Psab}{0}{\Mab}
\Vec{\gal\circ\Fab}{\gamma_{\alpha}\circ\Fab}
\label{TRANSSeq}\Eeq
and expansion in powers of $\zebe$ leads to four equations, the first of
which shows that the functions $p_{\alpha}$ satisfy
\Beq
p_{\beta}(\zbe) = p_{\alpha}(\Fab(\zbe)).
\Eeq
Using the fact that at level zero the functions $\Fab$ are simply the
transition functions of a Riemann surface, one finds, first at level
zero in the Grassmann generators, and then at higher levels by
induction, that
\Beq
p_{\alpha}=p
\label{PCONDeq}\Eeq
where $p$ is a constant element of $\Ceven$ which is independent of
$\alpha$. This allows
the other three equations derived from equation (\ref{TRANSSeq}) to be
simplified to give two independent conditions,
\Beqa
\sigma_{\beta} &=& (\sigma_{\alpha}\circ \Fab)\sqrt{\Fab'+
\Psab \Psab' } +
\Psab(r'_{\alpha}\circ \Fab)\sqrt{\Fab'}\label{ODDeq}\\
r'_{\beta} &=&
(r'_{\alpha}\circ\Fab)\Fab' +  \Psab' (\sigma_{\alpha}\circ\Fab)
 + \Psab (\sigma_{\alpha}'\circ\Fab)\Fab'\End
&&\label{EVENeq}
\Eeqa
which, with the condition (\ref{PCONDeq}), are necessary and sufficient
for $\Vecal$ to be the local representative a superconformal function.

The remainder of the proof involves constructing a linear map $\Phi$
from the
space $(\Ceven \otimes \Hom) \oplus (\Codd \otimes \Spim)$ into the
space of superconformal functions on $M$, and showing
that it is invertible. A typical element of the domain of $\Phi$ can be
expressed
as $p \oplus \pi$ with $p =\Bod{p}\One+p\Ind{12}\Ge1\Ge2+ \dots$ and
$\pi =\pi\Ind1\Ge1+\pi\Ind2\Ge2+ \dots$; however, since (\ref{ODDeq})
and (\ref{EVENeq}) are linear in $r$ and $\sigma$, it is sufficient to
consider
$\pi = \pi\Ind{i}\Ge{i}$ (with no summation), extending $\pi$ to the
full domain by
linearity. The map $\Phi$ will be constructed
inductively, the induction being over the number $n$ of generators
$\Ge1,\dots,\Ge{n}$ of the Grassmann algebra on which the supermanifold
is modelled. In order to establish that $\Phi(p \oplus \pi)$ is
superconformal as $n$ increases by $1$, it is necessary to use a further
induction, over the Grassmann level of the terms in equations
(\ref{ODDeq}) and
(\ref{EVENeq}).

It may be verified by direct calculation that the even equation
(\ref{EVENeq}) is satisfied at level zero and the odd equation
(\ref{ODDeq}) at level 1 if we set
\Beq
\Phi(p \oplus\pi)=
\Svec{p+\zeal\pi_{\alpha}\Ge{i}}{\pi_{\alpha}\Ge{i}}.
\Eeq
(so that $r'=0$ at level zero).

To satisfy (\ref{EVENeq}) at level $2$ we set
\Beq
\Phi(p\oplus\pi) =
\Svec{p \oplus \zeal\pi_{\alpha}(\zal)}{\pi_{\alpha}(\zal) + \zeal
\ral'(\zal)}
\Eeq
where the $\ral$ (which  are zero at level zero) are determined by the
following cohomological argument: we require
\Beq
\frac{d}{d\zbe}(\Psab(\zbe)\pi_{\alpha}(\Fab(\zbe))) = \rbe'(\zbe) -
\ral'(\Fab(\zbe))\Fab'(\zbe)
\label{ONEeq}\Eeq
at level $2$. To see that there exist unique $\ral$ (up to a constant)
such that this equation is satisfied  the relation (\ref{PSeq}) is used;
the $\Ge{j}$ component of this equation is
\Beq
\Psag\Ind{j} = \Psab\Ind{j}\circ \Fbbg + \sqrt{\Fbab'\circ\Fbbg}
\,\Psbg\Ind{j},
\Eeq
so that (using the fact that $\pi\Ind{i}$ is a spinor on $\Bod{M}$)
\Beqa
&&\qquad (\pi_{\alpha}\Ind{i}\circ\Fbag)\Psag\Ind{j} \End
&=&
(\pi_{\alpha}\Ind{i}\circ\Fbab\circ\Fbbg)(\Psab\Ind{j}\circ \Fbbg)
+
(\pi_{\beta}\Ind{i}\circ\Fbbg) \Psbg\Ind{j}.\End
\Eeqa
Thus the collection $\Seto (\pi_{\alpha}\Ind{i}\circ\Fbab)\Psab\Ind{j}|
\alpha,\beta \in \Lambda, \Val\cap\Vbe\not=\emptyset \Setc$ defines an
element of
$\Coh1$, (or, in looser phrasing which will be used in the remainder of
the paper, $(\pi_{\alpha}\Ind{i}\circ\Fbab)\Psab\Ind{j}$ is an element
of $\Coh1$). Now any class in $\Coh1$ has a constant representative
\cite{Gunnin}; thus there exist   $\ral\Ind{ij}$ (unique up to a
constant) in $\Cha0$ which satisfy (\ref{ONEeq}) at level $2$, and thus
(\ref{EVENeq}) is satisfied at level 2..

To satisfy (\ref{ODDeq}) at level $3$ we set
\Beq
\Phi(p\oplus\pi) =
\Vec{p+\zeal\sigma_{\alpha}(\zal)}{\sigma_{\alpha}(\zal)+\zeal\ral'
(\zal)}
\Eeq
with each component of $\ral=\ral\Ind{ij}\Ge{i}\Ge{j}$ determined as
before and $\sigma=\pi\Ind{j}\Ge{j} +
\sigma\Ind{ijk}\Ge{i}\Ge{j}\Ge{k}$ where $\sigma\Ind{ijk}$ is determined
from (\ref{ODDeq}) by cohomological methods which will now be described.

It is sufficient to verify the $\{ijk\}$ component of equation
(\ref{ODDeq}), assuming without loss of generality that $i <j <k$.
Thus it must be shown that
\Beqa
&&\qquad \sigbe\Ind{ijk} =
(\sigal\Ind{ijk}\circ\Bod{\Fab})\sqrt{\Bod{\Fab'}} \End
&+&
(\pial'{}\Ind{i}\circ\Bod{\Fab})\Fab\Ind{jk}{}\sqrt{\Bod{\Fab'}}
+\frac{\Half(\pial{}\Ind{i}\circ\Bod{\Fab})\Fab'\Ind{jk}{}}
{\sqrt{\Bod{\Fab'}}} \End
&+& \Lsa\Psab{}\Ind{j}(\ral'\Ind{ik}\circ\Bod{\Fab}\Rsa- \Lsa j \Lr
k\Rsa\End
&+& \Lsa\frac{(\pial\Ind{i}\circ\Bod{\Fab})\Psab\Ind{j}\Psab'\Ind{k}}{2
\sqrt{\Bod{\Fab'}}}\Rsa - \Lsa j \Lr k\Rsa.
\label{TCOMeq}\Eeqa

To show that $\sigal\Ind{ijk}$ can be chosen so that this
level
three equation is satisfied, first consider
\Beq
\TTab\Indd{ijk} = (\pial\Ind{i}\circ\Bod{\Fab})\sp2 \Fab\Ind{jk}.
\Eeq
Using the consistency condition \cite{CraRab}
\Beq
\Fag = \Fab \circ \Fbg + \Psbg (\Psab\circ\Bod{\Fbg})
\sqrt{\Fab'\circ\Fbg}
\Eeq
it then follows that
\Beq
\TTag\Indd{ijk} = \TTab\Indd{ijk} \circ \Bod{\Fbg} +
\TTbg\Indd{ijk} +
\Uabg\Indd{ijk}
\Eeq
where
\Beqa
\qquad\Uabg\Indd{ijk}  &=&
\Lsa(\pial\Ind{i}\circ\Bod{\Fag})\sp2 \Psbg\Ind{j}
(\Psab\Ind{k}\circ\Bod{\Fbg} )
\sqrt{\Bod{\Fab'}\circ\Bod{\Fbg}}\Rsa \End
&&\quad-\Lsa j \Lr k\Rsa.
\End
\Eeqa
Using (\ref{PSeq}) it may be shown that $\Uabg\Indd{ijk}$ is in
$\Coh2$. Also, it it may be verified by explicit calculation that
\Beq
\Diff{\Uabg\Indd{ijk}}{\ga} = -2(\Aag\Indd{ijk} -
(\Aab\Indd{ijk}\circ\Bod{\Fbg})\Bod{\Fbg'} -\Abg\Indd{ijk})
\Eeq
where
\Beqa
\Aab\Ind{ijk}
&=&\Lsa(\pial\Ind{i}\circ\Bod{\Fab})\Psab\Ind{j}(\ral'
\Ind{ik}\circ\Bod{\Fab})\Bod{\Fab'}\End
&&\qquad+
\Half(\pial\Ind{i}\circ\Bod{\Fab})\sp2\Psab\Ind{j}\Psab'\Ind{k}\Rsa
-\Lsa j \Lr k\Rsa.
\Eeqa
Thus, if functions $\Ffab$ are chosen on $\Val \cap \Vbe$ satisfying
\Beq
\frac{d\Ffab}{d\zbe} = -2\Aab\Indd{ijk},
\Eeq
then
\Beq
\frac{d}{d\zga}(\Uabg\Indd{ijk} -\Ffag+\Ffab\circ\Bod{\Fbg}+\Ffbg)=0
\Eeq
so that $\Uabg\Indd{ijk} -\Ffag+\Ffab\circ\Bod{\Fbg}+\Ffbg$ is an
element of $H\sp2(\Bod{M},\Comp)$. Now $H\sp2(\Bod{M},\Comp)$ is
trivial,
and so there exist $\Kab$ in $C\sp1(\Bod{M},\Comp)$ such that
\Beq
\Uabg\Indd{ijk} -\Ffag+\Ffab\circ\Bod{\Fbg}+\Ffbg = \Kag-
\Kab\circ\Bod{\Fbg} - \Kbg.
\Eeq
Thus, if $\Lab\Indd{ijk} = \Ffab + \Kab$,
\Beq
\Uabg\Indd{ijk} = \Lag\Indd{ijk} -\Lab\Indd{ijk}\circ\Bod{\Fbg} -
\Lbg\Indd{ijk},
\Eeq
and
\Beq
\Diff{\Lab\Indd{ijk}}{\be} = -2\Aab\Indd{ijk}.
\label{Leq}\Eeq
Thus $\TTab\Indd{ijk}-\Lab\Indd{ijk}$ is in $\Coh1$, and (again using
the
fact that every class in $\Coh1$ has a constant
representative), it can be deduced that there exist
$k_\al\Indd{ijk}$ in
$\Cha0$ such that
\Beq
\frac{d}{d\,\zbe}(\TTab\Indd{ijk}-\Lab\Indd{ijk} +
k_{\al}\Indd{ijk}\circ\Bod{\Fab} - k_{\be}\Indd{ijk} )=0.
\Eeq
Using (\ref{Leq}), we see that the level three equation (\ref{TCOMeq})
is then satisfied if
\Beq
\sigal\Ind{ijk} =
{\Diff{k_{\al}\Indd{ijk}}{\al}}/{2\pial\Ind{i}},
\Eeq
the division being possible because at any point where $\pial\Ind{i}$
has a zero, $k_\al\Indd{ijk}$ must have one of at least one order
higher.

The inductive construction of $\Phi(p \oplus \pi)$ is then completed by
similar arguments, although the details are too long to be included
here. Simple induction over level however is difficult because of cross
terms which arise; the double induction mentioned above, over number of
generators and level, avoids this difficulty.

The proof of the theorem is completed by showing that  $\Phi$ is
bijective. It is evident from its definition that $\Phi$ is injective.
To show that it is surjective, suppose that $G$ is a superconformal
function. From previous arguments we know that $G$ has local
representatives of the form
\Beq
\Vec {\gal(\zal,\zeal)}{\gamma_{\alpha}(\zal,\zeal)}= \Vec{p +
\zeal\sigal(\zal)}{\sigal(\zal) + \zeal\ral'(\zal)}.
\Eeq
Suppose that $\sigma$ is decomposed level by level as
\Beq
\sigma = \sigma\Lev1 + \sigma\Lev3 + \dots
\Eeq
where $\sigma\Lev{2k+1}$ denotes the component of $\sigma$ of level
$2k+1$ in the generators $\Ge1,\Ge2,\dots$ of $\Bare$.
Then, letting $\Sigt_{2k+1}$ be defined inductively by
\Beqa
\Sigt_1 &=& \sigma\Lev{1} \End
\Sigt_{2k+1} &=& \sigma\Lev{2k+1}  - \Lsa\Phi\Lba0 \oplus
(\sum_{\ell=1}\sp{k}
\Sigt_{2\ell-1})\Rba\Rsa,
\Eeqa
it follows from the construction of $\Phi$ that each
$\Sigt_{2k+1}$ is an element of $\Codd \otimes \Spim$, and also that
\Beq
G = \Phi\Lbb p \oplus (\sum_{k=0}\sp{\infty}\Sigt_{2k+1})\Rbb,
\Eeq
so that $\Phi$ must be surjective.
\hfil$\blacksquare$\break

Thus we have shown that the modified definition of a function embodied
in the notion of a superconformal function leads to a function space
with the desired properties. In the case of a split super Riemann
surface a superconformal function is simply a holomorphic function, but
the modification in the general case allows a constant structure as one
moves away from the split part of super moduli space. It seems likely
that this will
be useful in applications. Further developments to be considered are the
integration theory of such functions and the construction of analogous
fields of higher spin.
\vfil\eject

\end{document}